\documentclass[superscriptaddress,onecolumn]{revtex4}

\usepackage{graphicx}
\usepackage{amsmath}
\usepackage{amssymb}
\usepackage{multirow}
\usepackage{float}

\usepackage[dvipsnames]{xcolor}
\RequirePackage[colorlinks,citecolor=blue,urlcolor=blue,linkcolor=blue]{hyperref}

\newcommand{\be}{\begin{eqnarray}}
\newcommand{\ee}{\end{eqnarray}}
\newcommand{\bea}{\begin{eqnarray}}

\newcommand{\eea}{\end{eqnarray}}

\makeindex
\begin{document}

\title{Charged scalar field perturbations in Ernst black holes}

\author{Ram\'{o}n B\'{e}car}
\email{rbecar@uct.cl}
\affiliation{Departamento de Ciencias Matem\'{a}ticas y F\'{\i}sicas, Universidad Catolica de Temuco}

\author{P. A. Gonz\'{a}lez}
\email{pablo.gonzalez@udp.cl} \affiliation{Facultad de
Ingenier\'{i}a y Ciencias, Universidad Diego Portales, Avenida Ej\'{e}rcito
Libertador 441, Casilla 298-V, Santiago, Chile.}

\author{Yerko V\'{a}squez}
\email{yvasquez@userena.cl}
\affiliation{Departamento de F\'{\i}sica, Facultad de Ciencias, Universidad de La Serena,\\
Avenida Cisternas 1200, La Serena, Chile.}

\date{\today}

\begin{abstract}

We consider the propagation of a charged massive scalar field in the background of a four-dimensional Ernst black hole and study its stability analyzing the quasinormal modes (QNMs), which are calculated using the semi-analytical Wentzel-Kramers-Brillouin  method and numerically  using the continued fraction method. Mainly, we find that for 
a scalar field mass less than a critical mass, the decay rate of the QNMs decreases when the harmonic angular number $\ell$ increases; and for a scalar field mass greater than the critical mass the behaviour is inverted, i.e,  the longest-lived modes are always the ones with the lowest angular number recovering the standard behaviour. Also, we find a critical value of the external magnetic field, as well as, a critical value of the scalar field charge that exhibit the same behaviour with respect to the angular harmonic numbers.
In addition, we show that the spacetime allows stable quasibound states and we observe a splitting of the spectrum due to the Zeeman effect. Finally, we show that the unstable null geodesic in the equatorial plane is connected with the QNMs when the azimuthal quantum number satisfy $m= \pm \ell$ in the eikonal limit.

\end{abstract}

\maketitle

%\printindex

\tableofcontents

%\vspace{5cm}
\clearpage

%%%%%%%%%%%%%%%%%%%%%%%%%%
\section{Introduction}
%%%%%%%%%%%%%%%%%%%%%%%%%%

The black holes created in astrophysical processes are expected to be well described by asymptotically
flat solutions of Einstein equations. However, there is also a great interest in black holes with other kind
of asymptotic infinities. In particular, it is of interest to determine the effects that occur when black holes
are placed in an external background field, extending to infinity.
In this context, an exact solution of Einstein-Maxwell equations describing a black hole in a background magnetic universe was constructed about thirty years ago by Ernst \cite{Ernst:1976mzr}. In this model the external field is able of distorting the spherical symmetry of the geometry. The magnetic field has the effect of elongating the event horizon into a cigaret-shaped object with the long axis being parallel with the magnetic field lines. However, by astrophysical reasons the
magnetic field is supposed to be weak enough, so that, in this regime the
metric is well approximated by the Schwarzschild one, i.e. the magnetic field does not distort the geometry of the
space-time but only interacts with other electromagnetic
charges in the system.\\

Here, we study the propagation of charged massive scalar fields in Ernst black hole backgrounds. In this context, the quasinormal modes (QNMs) and quasinormal frequencies (QNFs)
\cite{Regge:1957a, Zerilli:1971wd, Zerilli:1970se, Kokkotas:1999bd, Nollert:1999ji, Berti:2009kk, Konoplya:2011qq, Aragon:2020qdc, Herrera:2017ztd} have recently acquired great interest due
to the detection of gravitational waves \cite{Abbott:2016blz}.
Despite the detected signal is consistent with the Einstein gravity \cite{TheLIGOScientific:2016src}, there are possibilities for alternative theories of gravity due to the large uncertainties in mass and angular momenta of the ringing black hole \cite{Konoplya:2016pmh}. Different investigations have emerged about QNMs of Ernst black hole, for example in \cite{Konoplya:2007yy} the authors studied massless scalar field perturbations and found that in the presence of a magnetic field the QNMs are longer lived and have larger oscillation frequencies, and in \cite{Kokkotas:2010zd,Turimov:2019afv} unstable modes in magnetized black holes were found. On the other hand, superradiant instability and the behaviour of the QNMs of a massive scalar field have been investigated in \cite{Brito:2014nja}  and \cite{Wu:2015fwa}. Their numerical results show that increasing the field effective mass and the magnetic field strength $B$ gives rise to a decreasing of the imaginary part of the QNMs until reaching a vanishing damping rate.
 \\

One of the aim of this work 
is to study the effect of the external magnetic field on the anomalous decay rate of the QNMs by using the six-order Wentzel-Kramers-Brillouin (WKB) method. It was shown that the imaginary part of the photon sphere QNFs has an anomalous behaviour for a scalar field mass less than a critical mass, i.e., the decay rate of the QNMs decreases when the harmonic angular number $\ell$ increases; and for a scalar field mass greater than the critical mass the behaviour is inverted, i.e,  the longest-lived modes are always the ones with lowest angular number recovering the standard behaviour. The critical mass corresponds to the value of the scalar field mass where the behaviour of the decay rate of the QNMs is inverted and can be obtained from the condition $Im(\omega)_{\ell}=Im(\omega)_{\ell+1}$ in the {\it eikonal} limit, that is when $\ell \rightarrow \infty$, and this behaviour have been studied in different black hole geometries: Schwarzschild, Schwarzschild-(A)dS, Reissner-Nordstr\"om, black hole in $f(R)$ gravity, for scalar and dirac fields \cite{Lagos:2020oek, Aragon:2020tvq, Aragon:2020xtm, Aragon:2020teq, Fontana:2020syy, Gonzalez:2022upu} and Bronnikov-Ellis and Morris-Thorne wormhole geometries \cite{Gonzalez:2022ote}. Also, it was shown the existence of a critical scalar field charge for Reissner-Nordstr\"om dS black hole \cite{Gonzalez:2022upu}.  Here, we show that there is a critical scalar field mass. Furthermore, there is a critical external magnetic field that exhibits the same behaviour with respect to the angular harmonic numbers, as well as, a critical scalar field charge, for charged massive scalar fields in Ernst  black hole backgrounds.\\

Then, we study the spectrum of quasibound states (QBS)
%is present \cite{Lasenby:2002mc} 
in this background by using the continued fraction method (CFM). QBS are localised in the black hole potential well and tend to zero at spatial infinity, and they have been studied over the years \cite {
%Damour:1976kh, Deruelle:1974zy, %Gaina:1993ib, Gaina:1992nx, 
GaltsovD:1983zpz, Lasenby:2002mc, Grain:2007gn, Sporea:2019iwk, Siqueira:2022tbc, Myung:2022krb}. 
Here, we show that the spectrum splits in $2 \ell+1$ branches, and the separation between the branches increases with the magnetic field, that is the analogous to the splitting of the energy levels of an atom in an external magnetic field, which is the well-known Zeeman effect. \\

Finally, we study the connection between the unstable null geodesics and the QNMs. It was shown that it occurs in Schwarzschild black holes \cite{Cardoso:2008bp}. However, such link
is violated in asymptotically flat black hole in the Einstein-Lovelock theory \cite{Konoplya:2017wot} and Schwarzschild AdS black holes \cite{Cardoso:2008bp}, see also \cite{Konoplya:2022gjp} for a further clarification. We will show that the unstable null geodesics in the equatorial plane are connected with the QNMs via the WKB method for the case $m = \pm \ell$ in the eikonal limit $\ell \rightarrow \infty $, 
for the four-dimensional Ernst  black hole. In order to see this phenomena for other spacetimes see \cite{Breton:2017hwe, Breton:2016mqh, Gallo:2015bda}, and references therein.\\

This work is organized as follows. In Sec.~\ref{background} we give a brief review of Ernst black holes. Then, in Sec.~\ref{CSP} we study the charged scalar field perturbations, and in Sec.~\ref{QNM} we calculate the QNFs by using the WKB method in order to study the anomalous decay rate for high values of $\ell$, and the CFM for small values of $\ell$. Then, we study the QBS in Sec.~\ref{QS}, and the unstable null geodesic in Sec. ~\ref{UNG}, in order to show if there is a link between the unstable null geodesics and the QNMs. Finally, we conclude in Sec.~\ref{conclusion}.

%%%%%%%%%%%%%%%%%%%%%%%%%%%%%%%%%%%%%%%%%%%%%
\section{Ernst black holes}
\label{background}
%%%%%%%%%%%%%%%%%%%%%%%%%%%%%%%%%%%%%%%%%%%%%%

The Ernst metric is an exact solution of the Einstein-Maxwell action. The Ernst solution can be interpreted as providing a model for the exterior spacetime due to a massive body which is placed in an external magnetic field, and have the feature of not being asymptotically flat. Its line element in Schwarzschild-like coordinate system is given by \cite{Ernst:1976mzr} 
\begin{equation}
  ds^{2}=\Lambda^2   \left( \left(1-\frac{2M}{r}\right)dt^2-\left(1-\frac{2M}{r}\right)^{-1}dr^2-r^2d\theta^2\right)
 -\frac{r^2\sin^2\theta}{\Lambda^2}d\phi^2\,, \label{Ernstmetric}
 \end{equation}
where 
\begin{equation}
    \Lambda=1+B^2r^2\sin^2\theta\,,
\end{equation}
and $B$ is the strength of the external magnetic field. The vector potential for the magnetic field is given by 
\begin{equation}
    A_{\mu}dx^{\mu}=-\frac{Br^2\sin^2\theta}{\Lambda}d\phi\,.
\end{equation}
As pointed out, in this model, the external field is able of distorting the spherical symmetry of the geometry. The magnetic field has the effect of elongating the event horizon into a cigaret-shaped object with the long axis being parallel with the magnetic field lines. The magnetic field lines remain perpendicular to all points on the event horizon, analogous to electric lines of force about a conductor. However, it was shown 
that the external magnetic field can be considered as a test field when the strength of the magnetic field
satisfies the condition $B<<B_M=\frac{c^{4}}{G^{3/2}}M_{\odot}(\frac{M_{\odot}}{M})\sim 10^{19}(\frac{M_{\odot}}{M})G$ \cite{Frolov:2010mi}. On the other hand, from an astrophysical point of view the magnetic field near the event horizon of stellar black holes $(10 M_{\odot})$ and supermassive black holes $(10^{9}\odot)$ are very small compared with $B_M$, thereby it is reasonable to neglect the distortions of curvature due to the external magnetic field around black holes. Note that the Schwarzschild black holes and the Melvin metric can be obtained when $B=0$, and $M=0$,  respectively. 

%%%%%%%%%%%%%%%%%%%%%%%%%%%%%%%%%%%%%%%%%%%%%%%%%%%55
\section{Charged scalar field perturbations}
\label{CSP}
%%%%%%%%%%%%%%%%%%%%%%%%%%%%%%%%%%%%%%%%%%%%%%%%%%%%%%%%5

A massive charged scalar field satisfies the Klein-Gordon equation,
\begin{equation}
\left(\nabla_{\alpha}+ i qA_{\alpha}\right)\left(\nabla^{\alpha}+ iqA^{\alpha}\right)\Psi+ \mu^2 \Psi=0 \,,
\end{equation}
where $\mu$ is the mass of the scalar field and $q$ its charge.
The problem can be reasonably simplified by making the following assumption: for small $B$, one can safely neglect terms higher than $B^{2}$, which allows to separate the radial and angular variables. In this manner, by taking into account the
spacetime symmetry, we can write $\Psi$ as
\begin{equation}
\Psi(t,r,\theta,\phi)=\frac{1}{r}R(r)S(\theta)e^{-i\omega t}e^{i m\phi}
\end{equation}
where $m$ is  the  azimuthal  quantum  number  and $\omega$ is  the
QNF of the mode. Thus, the Klein-Gordon equation reads
\begin{equation} \label{radial}
 \frac{1}{r}\frac{d}{dr}\left(r^{2}f(r)\frac{d}{dr}(\frac{R(r)}{r})\right)
 +\left(\frac{\omega^{2}}{f}-4B^{2}m^{2}+2mqB-\mu^{2}-\frac{\ell (\ell+1)}{r^{2}}\right)=0\,,
\end{equation}
where $\ell$ is the harmonic angular number and $f(r)= 1-2M/r$. Now, by using the tortoise coordinate $r^{\ast}$ given by $dr^{\ast}=\frac{dr}{f(r)}$, the Klein-Gordon equation can be written as a one-dimensional Schr\"odinger-like equation
\begin{equation}\label{schrodinger}
\frac{d^{2}R(r^{\ast})}{dr^{\ast2}}+(\omega^{2}-V_{eff}(r))R(r^{\ast}) =0\,,
\end{equation}
with an effective potential $V_{eff}(r)$ given by
\begin{equation}
V_{eff}(r)=f(r)\left(\frac{f'(r)}{r}+\frac{\ell(\ell+1)}{r^{2}}+4B^{2}m^{2}-2mqB+\mu^{2}\right)\,,
\end{equation}
whose asymptotic behaviours near the event horizon and at spatial infinity are
\begin{align*}
V_{eff}(r\rightarrow r_{h})&=0 \,,\\
V_{eff}(r\rightarrow \infty)=\mu_{eff}^{2}&=\mu^{2}-2mqB+4m^2B^{2}\,.
\end{align*}
Clearly, the potential coincides with the Schwarzschild potential of a massive scalar field when its mass is replaced by the effective mass
\begin{equation}
\mu_{eff}^{2}=4 m^{2} B^{2}-2mqB+\mu^{2}\,.
\end{equation}

\begin{figure}[ht]  
\begin{center}
\includegraphics[width=0.6\textwidth]{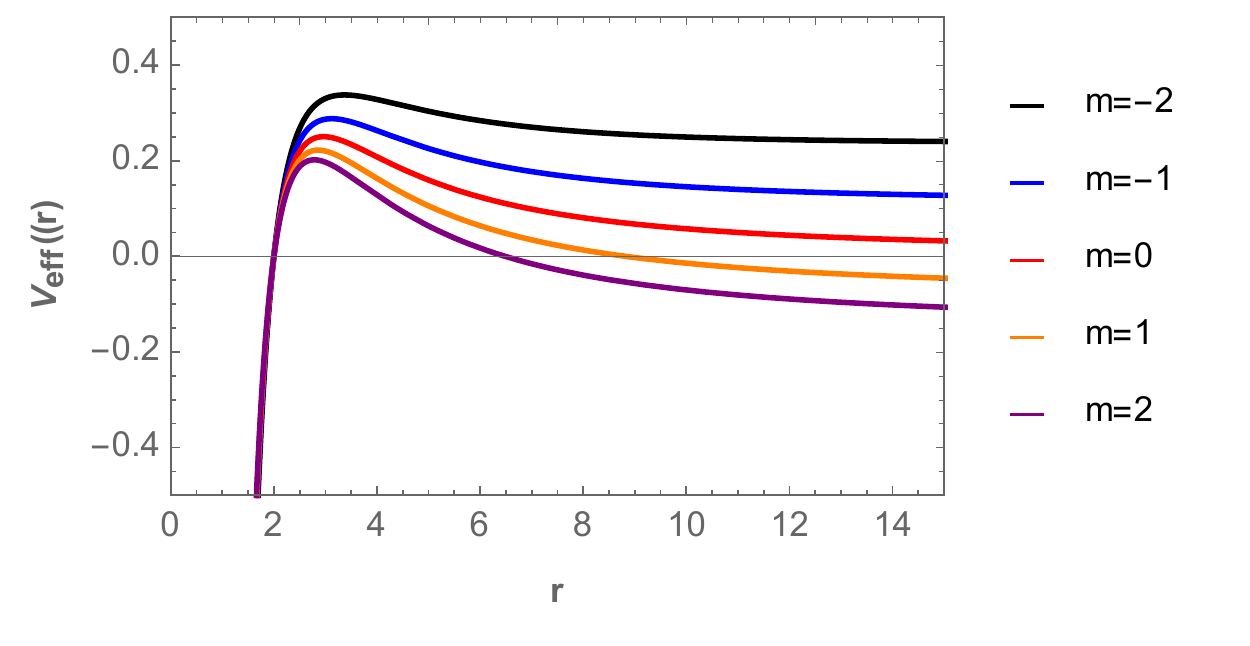}
\caption{Effective potential for the multipole number $\ell=2$ with $B=0.05$, $q=1$, $M=1$ and $\mu=0.1$}
\label{potentialeff}
\end{center}
\end{figure}

Note that the value of the squared effective mass can be negative, depending on the values of mass and charge of the scalar field, azimuthal number, and strength of the magnetic field, which is associated with the existence of unstable modes \cite{Kokkotas:2010zd,Turimov:2019afv}. In Fig. \ref{potentialeff} the radial dependence of the effective potential is illustrated, where it is possible to observe a potential barrier. In section \ref{QS} we will show that the potential also can have the shape of a potential well for some values of the parameters. Also, as was pointed out, if the value of the squared effective mass is positive, there is some threshold value of the effective mass after that the effective potential loses its barrier-like form and the QNMs disappear, beyond this value, there are arbitrarily long lived QNMs, so called quasi-resonance modes \cite{Ohashi:2004wr, Toshmatov:2016bsb, Toshmatov:2017qrq}.

%%%%%%%%%%%%%%%%%%%%%%%%%%%%%%%%%%%%%%%%%%%%%5%%
\section{Quasinormal modes}
\label{QNM}
%%%%%%%%%%%%%%%%%%%%%%%%%%%%%%%%%%%%%%%%%%%%%%%%

In this section we calculate the QNFs of the scalar field using the semi-analytical WKB method and numerically using the CFM. We study the anomalous decay rate of the QNMs and also the splitting of the spectrum due to the external magnetic field.

\subsection{QNMs using WKB method}

In order to get some analytical insight about the behaviour
of the QNFs, we use the WKB method \cite{Mashhoon, Schutz:1985zz, Iyer:1986np, Konoplya:2003ii, Matyjasek:2017psv, Konoplya:2019hlu} that can be used for effective potentials which have the form of a potential barrier, approaching to a constant value at the event horizon and cosmological horizon or spatial infinity \cite{Konoplya:2011qq}. Here, we consider the eikonal limit $\ell \rightarrow \infty$   to estimate the critical scalar field mass, by
considering $\omega_I^\ell=\omega_I^{\ell+1}$ as a proxy for where the transition or critical behaviour occurs \cite{Lagos:2020oek}.
The QNMs that belongs to the photon sphere family are determined by the behaviour of the effective potential near its maximum value, located at the position $r^*_{max}$. The Taylor series expansion of the potential around its maximum is given by
\begin{equation}
V(r^*)= V(r^*_{max})+ \sum_{i=2}^{\infty} \frac{V^{(i)}}{i!} (r^*-r^*_{max})^{i}   \,,
\end{equation}
where
\begin{equation}
V^{(i)}= \frac{d^{i}}{d r^{*i}}V(r^*) \Big|_{r^*=r^*_{max}}
\end{equation}
corresponds to the $i$-th derivative of the potential with respect to the torsoise coordinate $r^*$ evaluated at the position of the maximum $r^*_{max}$. Using the WKB approach carried to third order beyond the eikonal approximation it was found that the QNFs are given by, see for instance \cite{Hatsuda:2019eoj}
\begin{equation} \label{omega}
\omega^2 = V(r^*_{max}) -2 i U \,,
\end{equation}
where
\begin{multline}
 U = N\sqrt{-V^{(2)}/2}
+\frac{i}{64} \left( -\frac{1}{9}\frac{V^{(3)2}}{V^{(2)2}} (7+60N^2)+\frac{V^{(4)}}{V^{(2)}}(1+4 N^2) \right)+\frac{N}{2^{3/2} 288} \Bigg( \frac{5}{24} \frac{V^{(3)4}}{(-V^{(2)})^{9/2}} (77+188N^2) +\\
\frac{3}{4} \frac{V^{(3)2} V^{(4)}}{(-V^{(2)})^{7/2}}(51+100N^2)
+ \frac{1}{8} \frac{V^{(4)2}}{(-V^{(2)})^{5/2}}(67+68 N^2)
+\frac{V^{(3)}V^{(5)}}{(-V^{(2)})^{5/2}}(19+28N^2)+\frac{V^{(6)}}{(-V^{(2)})^{3/2}} (5+4N^2)  \Bigg)\,,
\end{multline}
and $N=n+1/2$, with $n=0,1,2,\dots$, the overtone number.
\newline

Defining $L^2= \ell (\ell+1)$, we find that for large values of $L$, the maximum of the potential is located approximately at
\begin{equation} \label{radius}
r_{max} \approx r_0+ \frac{1}{L^2} r_1 + \mathcal{O}(L^{-4})~,
\end{equation}
where
\begin{eqnarray}
 \nonumber r_0 &=& 3M \,,\\
 r_1 &=& (-1+108m^2 B^2 M^2 -54 m B M^2 q +27 \mu^2 M^2) \frac{M}{3}\,,
\end{eqnarray}
and
\begin{equation} \label{coa}
V(r^*_{max}) \approx \frac{1}{27 M^2} L^2 +\frac{2+108 m^2 B^2 M^2 -54 m q B M^2 +27 \mu^2 M^2}{81 M^2} +\mathcal{O}(L^{-2}) \,,
\end{equation}
the second derivative of the potential evaluated at $r^*_{max}$ is given by
\begin{equation}
V^{(2)}(r^*_{max}) \approx -\frac{2}{729 M^4} L^2 +\frac{4 (-4+108 m^2 B^2 M^2 -54 m q B M^2 +27 \mu^2 M^2)}{6561 M^4} +\mathcal{O}(L^{-2}) \,,
\end{equation}
and the higher derivatives of the potential evaluated at $r^*_{max}$ yield the following expressions\\
\begin{eqnarray}
 \nonumber V^{(3)}(r^*_{max}) &\approx&   \frac{4}{6561 M^5} L^2 + \mathcal{O}(L^0)  \,, \\
 \nonumber
 V^{(4)}(r^*_{max})  &\approx& \frac{16}{19683 M^6} L^2 + \mathcal{O}(L^0) \,, \\
 \nonumber
 V^{(5)}(r^*_{max}) &\approx& -\frac{40}{59049 M^7} L^2 + \mathcal{O}(L^0) \,,  \\
 \nonumber
  V^{(6)}(r^*_{max}) &\approx& -\frac{64}{177147 M^8} L^2 + \mathcal{O}(L^0) \,.
\end{eqnarray}
Using these results together with Eq. (\ref{omega}) we obtain:
\begin{eqnarray}
\label{omegarelation}
\nonumber \omega &\approx& \frac{1}{3 \sqrt{3} M} L - \frac{1+2n}{6 \sqrt{3} M} i + \frac{17-15n (n+1)+1944 m^2 B^2 M^2 - 972 m q B M^2}{324 \sqrt{3} M} L^{-1}  \\
 && - \frac{(1+2n)(137+235 n (n+1) -116640 m^2 B^2 M^2 + 58320 m q B M^2-29160M^{2}\mu^{2})}{23328 \sqrt{3} M} i L^{-2} + \mathcal{O} (L^{-3})\,.
\end{eqnarray}

The term of order $L^{-2}$ vanishes at the value of the critical mass $\mu_c$, which is given by \\
\begin{equation}
\label{mc}
%\nonumber
\mu_c M  =  \frac{1}{54} \sqrt{\frac{137}{10} + \frac{47}{2} n (n+1)-11664 m^2 B^2 M^2 + 5832 m q B M^2}
\,.
\end{equation}
For $B=0$ the critical mass of the scalar field in a Schwarzschild background is recovered \cite{Lagos:2020oek}.
Also, it is possible to obtain a critical value of the magnetic field, even for zero scalar field mass $\mu$, which is given by
\begin{equation}
\label{Bc}
M B_c = \frac{270 m q M + \sqrt{10 m^2 (137+235 n (n+1)+ 7290 M^2 (q^2-4\mu^2))}}{1080 m^2}\,,
\end{equation}
for $\mu=0$, $q=0$, and $m \neq 0$ this yields $M B_c = \frac{1}{108 m} \sqrt{\frac{137}{10} + \frac{47}{2} n (n+1)}$. Also, there is a critical charge for the scalar field given by
%\newline
\begin{equation}
\label{qc}
 M q_c = \frac{-137 - 235 n(n+1)+29160 M^2(\mu^2 + 4 m^2 B^2)}{58320 m M B}  \,.
\end{equation}

\begin{figure}[ht] 
\begin{center}
\includegraphics[width=0.5\textwidth]{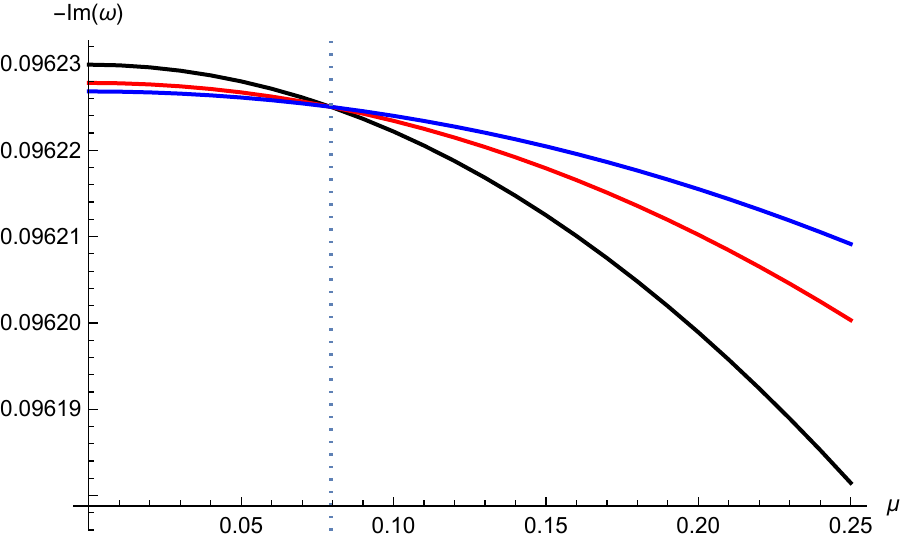}
\caption{The behaviour of
$-Im(\omega)$ 
as a function of $\mu$, with $n=0$, $M=1$, $B=0.01$, $q=0.1$ and $m=1$, using the six-order WKB method.
Here, the WKB method gives via Eq. (\ref{mc}) $\mu_c \sim 0.0794$, vertical dotted line. Black line for $\ell=30$, red line for $\ell=40$, and blue line for $\ell=50$.}
\label{FanomalousA}
\end{center} 
\end{figure}

\begin{figure}[ht] 
\begin{center}
\includegraphics[width=0.5\textwidth]{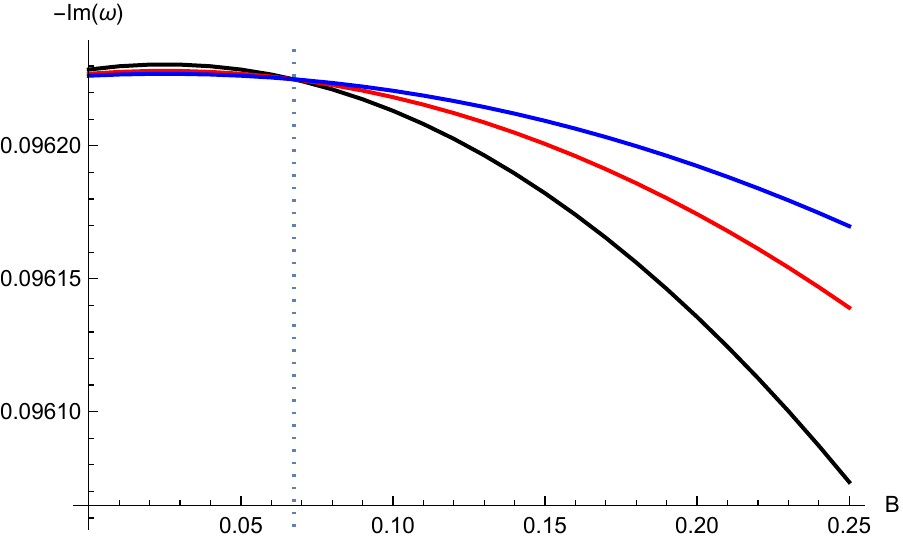}
\caption{The behaviour of
$-Im(\omega)$ 
as a function of $B$, with $n=0$, $M=1$, $q=0.1$, $m=1$, and $\mu=0$, using the six-order WKB method.
Here, the WKB method gives via Eq. (\ref{Bc}) $B_c \sim 0.0674$ , vertical dotted line. Black line for $\ell=30$, red line for $\ell=40$, and blue line for $\ell=50$.}
\label{FanomalousB}
\end{center}
\end{figure}

\begin{figure}[ht] 
\begin{center}
\includegraphics[width=0.5\textwidth]{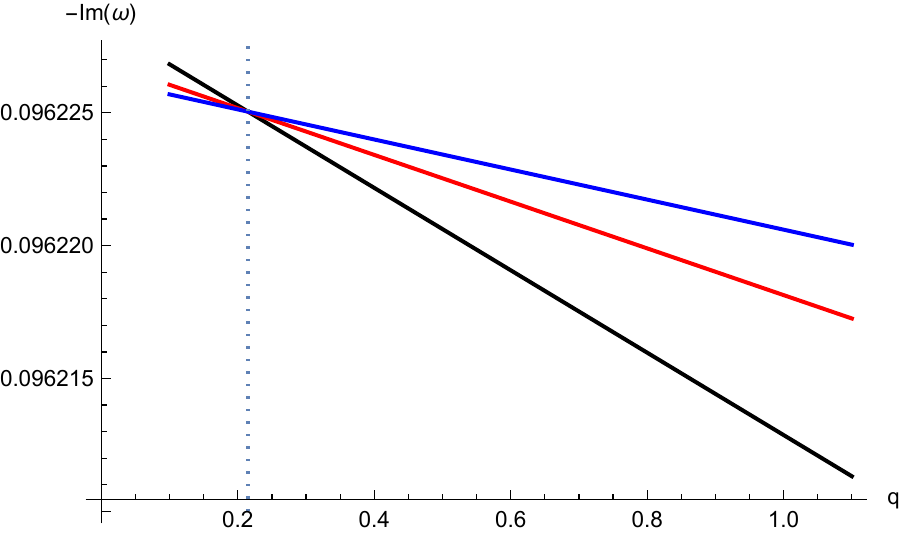}
\caption{The behaviour of
$-Im(\omega)$ 
as a function of $q$, with $n=0$, $M=1$, $B=0.01$, $m=-1$, and $\mu=0$, using the six-order WKB method.
Here, the WKB method gives via Eq. (\ref{qc}) $q_c \sim 0.215$ , vertical dotted line. Black line for $\ell=30$, red line for $\ell=40$, and blue line for $\ell=50$.}
\label{Fanomalousq}
\end{center}
\end{figure}

In Fig. \ref{FanomalousA} we show the behaviour of $-Im(\omega)$ as a function of the scalar field mass $\mu$. We can observe a critical scalar field mass, where for small values of the scalar field mass the longest-lived mode is the mode with highest angular number $\ell$, while that for values of the scalar field mass greater that the critical one the longest-lived mode is the mode with smallest angular number. Also, in Fig. \ref{FanomalousB} we can observe a similar behaviour for external magnetic field and for the charge of the scalar field in Fig. \ref{Fanomalousq}. 

\newpage

\subsection{QNMs using the CFM}

In this section we investigate the behaviour of the QNMs of charged massive scalar fields around the Ernst black hole using the CFM, devised by Leaver to compute the QNMs of Schwarzschild and Kerr black holes \cite{Leaver:1985ax, Leaver:1990zz}, and improved later by Nollert \cite{Nollert:1993zz}. It has been used to computed the QNMs in several situations and in particular for charged fields around charged black holes in \cite{Konoplya:2013rxa, Richartz:2014jla, Chowdhury:2018pre}.
\newline

The boundary conditions are given by ingoing waves at the event horizon and outgoing waves at spatial infinity. Thus, 
\begin{equation}
R(r) \approx \left\{ { (r-r_h)^{- i  \omega r_h} \,\,\,\,\,\,\,\,\, \text{as}\,\,\,\,\, r \rightarrow r_h  \atop e^{i \Omega r} r^{i r_h (\Omega^2 + \omega^2) /2 \Omega} \,\,\,\,\, \text{as} \,\,\,\,\, r \rightarrow \infty} \right.
\end{equation}
where $\Omega = \sqrt{\omega^2 - \mu^2 + 2 m q B - 4 m^2 B^2}$. Now, considering the following ansatz for the solution to the radial equation (\ref{radial}) which incorporates the desired boundary conditions
\begin{equation}
    R(r) = (r-r_h)^{- i\omega r_h} e^{i \Omega r} r^{i r_h (\Omega^2 + \omega^2)/2 \Omega + i \omega r_h} \sum_{n=0}^{\infty} a_n \left( \frac{r-r_h}{r} \right)^n\,,
\end{equation}
and substituting it in Eq. (\ref{radial}) a three term-recurrence relation is obtained for the coefficients
\begin{eqnarray}
\alpha_0 a_1 + \beta_0 \nonumber a_0 &=& 0  \,,   \\
\nonumber   \alpha_n a_{n+1} + \beta_n a_n + \gamma_n a_{n-1} &=& 0\,,
\end{eqnarray}
where
\begin{eqnarray}
\notag \alpha_n &=& -4 \Omega^2 (1+n)(1+n -2 r_h i \omega) \,, \\
\notag \beta_n &=& -2 \Omega \bigg( -2 \Omega (1+ \ell (\ell+1)+ 2n (n+1)) + (1+2n) i r_h (\omega+\Omega) (\omega +3\Omega) + 2 r_h^2 (\omega+\Omega)^3 \bigg) \,, \\
\gamma_n &=& - (2n \Omega - i r_h (\omega+ \Omega)^2)^2 \,.
\end{eqnarray}
The recursion coefficients must satisfy the following continued fraction relation for the convergence of the series
\begin{equation}
\beta_0 - \frac{\alpha_0 \gamma_1}{\beta_1-}\frac{\alpha_1 \gamma_2}{\beta_2-} \cdots\frac{\alpha_n \gamma_{n+1}}{\beta_{n+1}-} \cdots = 0\,.
\end{equation}
and the continued fraction must be truncated at some large index $N$. The QNFs are obtained solving this equation numerically.
\newline

\begin{figure}[ht] 
\begin{center}
\includegraphics[width=0.4\textwidth]{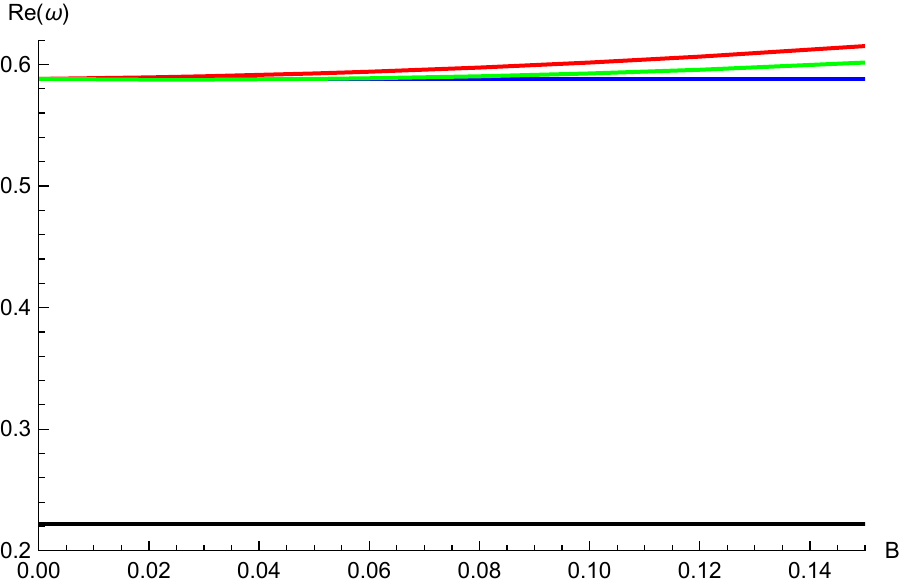}
\includegraphics[width=0.4\textwidth]{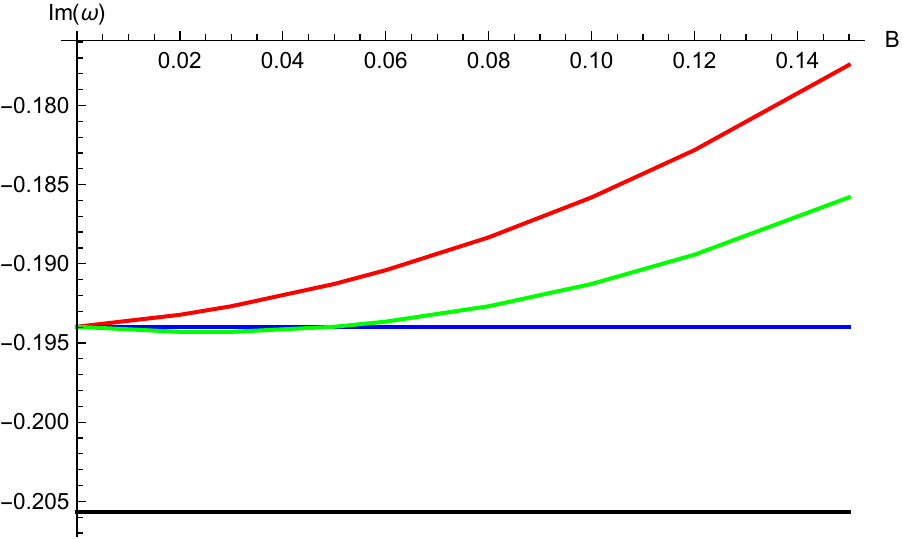}
\caption{Real and imaginary parts of fundamental QNFs as a function of $B$ for $\mu =0.1$, $r_h=1$, $q=0.1$ using the CFM. Black line for $\ell=m=0$, red line for $\ell=1$ and $m=-1$, blue line for $\ell=1$ and $m=0$, and green line for $\ell=1$ and $m=1$.}
\label{Fig1}
\end{center}
\end{figure}

In Table \ref{Table1} we show the QNFs for $\mu =0.1$, $r_h=1$, and $q=0.1$. The results show the splitting of the spectrum of the QNFs due to the Zeeman effect which increases with the magnetic field, and in Fig. \ref{Fig1} we plot the different branches.
Note that, for $m=-1$ (red line), the real oscillation frequency grows, and the damping rate is decreasing when the magnetic field is increasing, whereas for $m=1$ (green line) the real oscillation frequency decreases very slightly and then begin to grow, and the damping rate increases and then decreases when the magnetic field increases. On the other hand, for $m=0$ we recover the QNFs of Schwarzschild \cite{Konoplya:2004wg}, because the effective mass reduces to the mass of the scalar field.

\begin{table}[ht]
\caption{Fundamental QNFs for $\mu =0.1$, $r_h=1$, $q=0.1$, and different values of the angular number and $B$.
}
\label{Table1}\centering
\resizebox{\columnwidth}{!}{
\begin{tabular}{ | c | c | c | c | c | c | c | c |  }
\hline
$\ell$ & $m$ & $B=0$ & $B=0.02$ & $B= 0.05$ & $B=0.1$ & $B=0.15$  \\ \hline
 $0$  & $0$ & 0.22198957 - 0.20569164 i &  0.22198957 - 0.20569164 i &  0.22198957 - 0.20569164 i &  0.22198957 - 0.20569164 i &  0.22198957 - 0.20569164 i
\\\hline
\multirow{3}{0.2cm}{$1$} & $-1$ & 0.58810863 - 0.19397596 i & 0.58936199 - 0.19322116 i & 0.59258818 - 0.19127311 i &  0.60157396 - 0.18580627 i   & 0.61511616 - 0.17744451 i \\ 
 & $0$ & 0.58810863 - 0.19397596 i & 0.58810863 - 0.19397596 i & 0.58810863 - 0.19397596 i & 0.58810863 - 0.19397596 i &  0.58810863 - 0.19397596 i  \\
 & $1$ & 0.58810863 - 0.19397596 i & 0.58757170 - 0.19429897 i & 0.58810863 - 0.19397596 i & 0.59258818 - 0.19127311 i &  0.60157396 - 0.18580627 i \\ \hline
\multirow{5}{0.2cm}{$2$} & $-2$ & 0.96886635 - 0.19297645 i & 0.97114110 - 0.19219610 i & 0.97835671 - 0.18971583 i &  1.00062786 - 0.18201005 i    &  1.03599688 - 0.16959707 i  \\ 
 & $-1$ & 0.96886635 - 0.19297645 i & 0.96975076 - 0.19267314 i & 0.97202623 - 0.19189226 i & 0.97835671 - 0.18971583 i & 0.98787950 - 0.18643054 i \\
 & $0$ & 0.96886635 - 0.19297645 i & 0.96886635 - 0.19297645 i & 0.96886635 - 0.19297645 i & 0.96886635 - 0.19297645 i & 0.96886635 - 0.19297645 i   \\
 & $1$ & 0.96886635 - 0.19297645 i & 0.96848740 - 0.19310637 i & 0.96886635 - 0.19297645 i & 0.97202623 - 0.19189226 i & 0.97835671 - 0.18971583 i \\
 & $2$ & 0.96886635 - 0.19297645 i & 0.96861371 - 0.19306307 i & 0.97202623 - 0.19189226 i &  0.98787950 - 0.18643054 i  &  1.01664735 - 0.17641642 i \\ \hline
\end{tabular}
}
\end{table}

\section{Quasibound states}
\label{QS}

\begin{figure}[ht]  
\begin{center}
\includegraphics[width=0.45\textwidth]{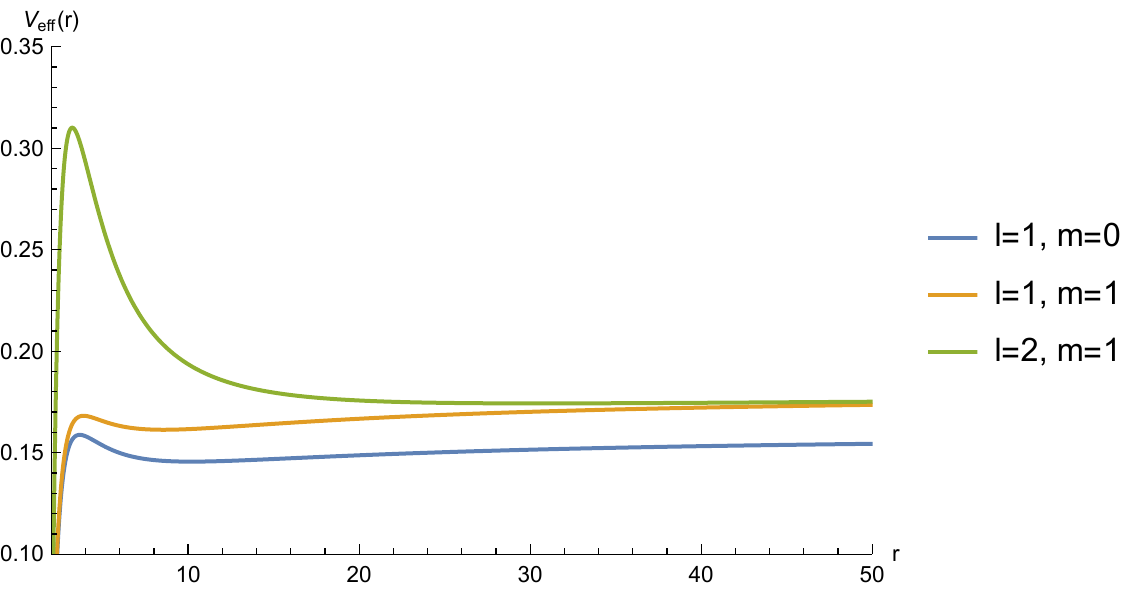}
\caption{Effective potential as a function of $r$ for $r_h = 2$, $\mu = 0.4$, $B = 0.1$, $q = 0.1$, and different values of $\ell$ and $m$.}
\label{FigQS}
\end{center}
\end{figure}

In this section we investigate the behaviour of the QBS of massive scalar fields around the Ernst black hole using the CFM. The computations of QBS and QNMs are very similar; however, the boundary conditions are different, for the QBS we must consider ingoing waves at the event horizon and evanescent waves at spatial infinity.
In Fig. \ref{FigQS} we show the effective potential for $r_h = 2$, $\mu = 0.4$, $B = 0.1$, $q = 0.1$, and different values of $\ell$ and $m$. Note that for some values of the parameters the effective potential allows potential wells, therefore QBS eventually can appear.

Now, the boundary conditions are given by 
%\newpage
\begin{equation}
R(r) \approx  \left\{ { (r-r_h)^{- i  \omega r_h} \,\,\,\,\, \text{as}\,\,\,\,\, r \rightarrow r_h \atop e^{- \Omega r} r^{- r_h (\Omega^2 - \omega^2) /2 \Omega} \,\,\,\,\, \text{as} \,\,\,\,\, r \rightarrow \infty} \right.
\end{equation}
where $\Omega = \sqrt{ \mu^2 -\omega^2 - 2 m q B + 4 m^2 B^2}$. So, considering the following ansatz for the radial function $R(r)$ which incorporates the desired boundary conditions
\begin{equation}
    R(r) = (r-r_h)^{- i\omega r_h} e^{- \Omega r} r^{-r_h (\Omega^2 - \omega^2)/2 \Omega + i \omega r_h} \sum_{n=0}^{\infty} a_n \left( \frac{r-r_h}{r} \right)^n\,,
\end{equation}
and substituting it in Eq. (\ref{radial}) we obtain a three term recurrence relation for the coefficients
\begin{eqnarray}
\alpha_0 a_1 + \beta_0 \nonumber a_0 &=& 0 \,,  \\
\nonumber   \alpha_n a_{n+1} + \beta_n a_n + \gamma_n a_{n-1} &=& 0 \,,
\end{eqnarray}
where
\begin{eqnarray}
\notag \alpha_n &=& 4 \Omega^2 (1+n)(1+n -2 r_h i \omega) \,, \\
\notag \beta_n &=& 2 \Omega \bigg( -2 \Omega (1+ \ell (\ell+1)+ 2n (n+1)) + (1+2n) r_h (\omega+ i \Omega) (\omega +3 i \Omega) + 2 r_h^2 (i \omega-\Omega)^3 \bigg) \,, \\
\gamma_n &=& (2n \Omega + r_h (\Omega -i  \omega)^2)^2 \,.
\end{eqnarray}

The recursion coefficients must satisfy the following continued fraction relation for the convergence of the series
\begin{equation}
\beta_0 - \frac{\alpha_0 \gamma_1}{\beta_1-}\frac{\alpha_1 \gamma_2}{\beta_2-} \cdots\frac{\alpha_n \gamma_{n+1}}{\beta_{n+1}-} \cdots = 0\,.
\end{equation}
and the continued fraction must be truncated at some large index $N$. The frequencies are obtained solving this equation numerically.
\newline

\begin{figure}[ht] 
\begin{center}
\includegraphics[width=0.4\textwidth]{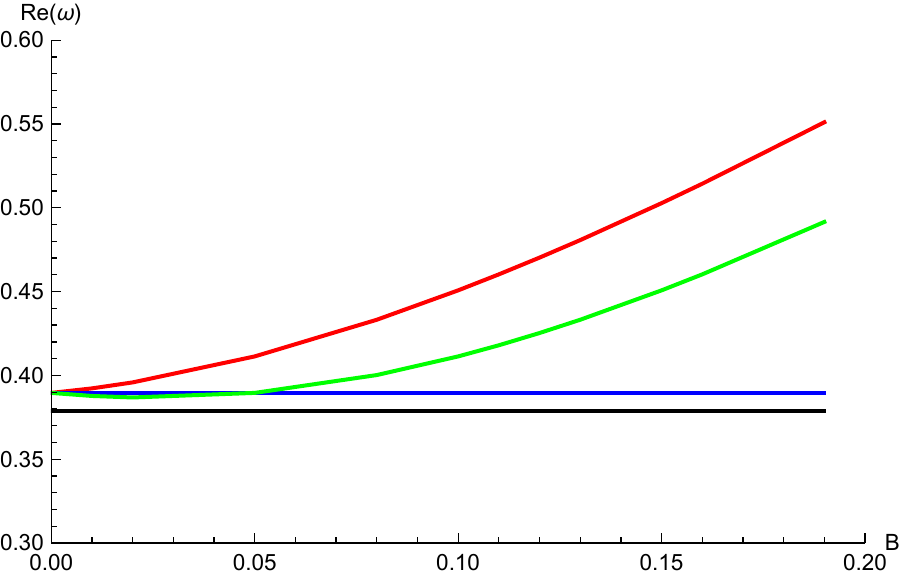}
\includegraphics[width=0.4\textwidth]{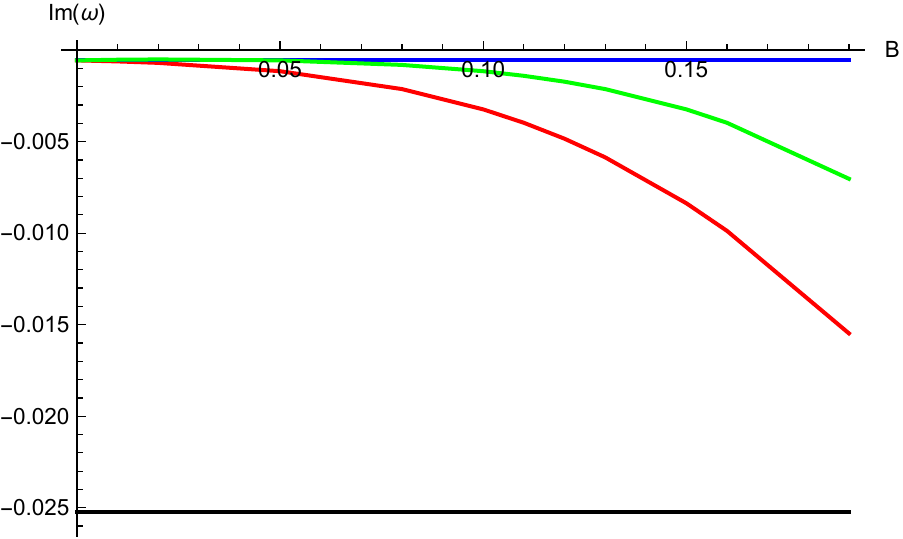}
\caption{Real and imaginary parts of fundamental frequencies of the QBS as a function of $B$ for $\mu =0.4$, $r_h=2$, $q=0.1$ using the CFM. Black line for $\ell=m=0$, red line for $\ell=1$ and $m=-1$, blue line for $\ell=1$ and $m=0$, and green line for $\ell=1$ and $m=1$.}
\label{Fig2}
\end{center}
\end{figure}

In Table \ref{Table2} we show the fundamental frequencies of the QBS for $\mu =0.4$, $r_h=2$, $q=0.1$, and different values of $\ell$, $m$ and $B$. The results show the splitting of the spectrum of the QBS due to the Zeeman effect, which increases with the magnetic field. In Fig. \ref{Fig2} we plot the different branches. Also,  we observe that the effective mass of the scalar field is slightly larger than the real part of the frequency, and the real part is much bigger than the imaginary part which are typical characteristics of QBS. The imaginary part is negative and therefore the modes are stable. 
Also, note that for $m = -1$ (red line), the real oscillation frequency grows, and the damping rate is increasing when the magnetic field is increasing, whereas for $m=1$ (green line) the real oscillation frequency decreases very slightly and then begin to grow, and the damping rate increases and then decreases when the magnetic field  increases. So, the real oscillation frequency of the QNMs and QBS has a similar behaviour, whereas for the damping rate the behaviour is opposite, when the magnetic field is increasing. On the other hand, for $m=0$ we recover the QBS of Schwarzschild \cite{Grain:2007gn}, because the effective mass reduces to the mass of the scalar field.

\begin{table}[ht]
\caption{Fundamental frequency of the QBS for $\mu =0.4$, $r_h=2$, $q=0.1$, and different values of the angular number and $B$.
}
\label{Table2}\centering
\resizebox{\columnwidth}{!}{
\begin{tabular}{ | c | c | c | c | c | c | c | c | c |  }
\hline
$\ell$ & $m$ & $B=0$ & $B=0.01$ & $B=0.02$ & $B= 0.05$ & $B=0.08$  &$B=0.1$  \\ \hline
 $0$  & $0$ & 0.37890623 - 0.02524730 i &  0.37890623 - 0.02524730 i  & 0.37890623 - 0.02524730 i  & 0.37890623 - 0.02524730 i &  0.37890623 - 0.02524730 i & 0.37890623 - 0.02524730 i   
\\\hline
\multirow{3}{0.2cm}{$1$} & $-1$ & 0.38955566 - 0.00056274 i  & 0.39225614 - 0.00061886 i & 0.39581751 - 0.00069974 i &  0.41132498 - 0.00115433  i &  0.43319420 - 0.00213047 i  &   0.45072920 - 0.00323952 i    \\ 
 & $0$ & 0.38955566 - 0.00056274 i  & 0.38955566 - 0.00056274 i  & 0.38955566 - 0.00056274 i & 0.38955566 - 0.00056274 i & 0.38955566 - 0.00056274 i  & 0.38955566 - 0.00056274 i  \\
 & $1$ & 0.38955566 - 0.00056274 i  &  0.38774097 - 0.00052743 i & 0.38682924 - 0.00051039 i &  0.38955566 - 0.00056274 i &  0.40020816 - 0.00081087 i   & 0.41132498 - 0.00115433 i    \\
 \hline
$\ell$ & $m$ & $B=0.11$ & $B=0.12$ & $B= 0.13$ & $B=0.15$ & $B=0.16$ & $B=0.19$    \\ \hline
 $0$  & $0$ &  0.37890623 - 0.02524730 i  & 0.37890623 - 0.02524730 i  & 0.37890623 - 0.02524730 i  & 0.37890623 - 0.02524730 i   &   0.37890623 - 0.02524730 i &  0.37890623 - 0.02524730 i
\\\hline
\multirow{3}{0.2cm}{$1$} & $-1$ & 0.46026339 - 0.00397261 i   & 0.47026206 - 0.00484167 i    &  0.48069416 - 0.00585783 i   &   0.50274895 - 0.00836835 i  &   0.51432342 - 0.00987717 i  &  0.55099232 - 0.01547377 i  \\ 
 & $0$ &  0.38955566 - 0.00056274 i  & 0.38955566 - 0.00056274 i  &   0.38955566 - 0.00056274 i  &  0.38955566 - 0.00056274 i &  0.38955566 - 0.00056274 i   &  0.38955566 - 0.00056274 i \\
 & $1$ & 0.41796647 - 0.00140621 i &  0.42527093 - 0.00172718 i &   0.43319420 - 0.00213047 i  &  0.45072920 - 0.00323952 i  &   0.46026339 - 0.00397261 i    &  0.49153157 - 0.00703072 i   \\ \hline

\end{tabular}
}
\end{table}

%%%%%%%%%%%%%%%%%%%%%%%%%%%%%%%%%%%%%%%%%%%%%%%%%%%5

%%%%%%%%%%%%%%%%%%%%%%%%%%%%%%%%%%%%%55

%\clearpage
%\newpage

\section{Unstable null geodesics}
\label{UNG}
%%%%%%%%%%%%%%%%%%%%%%%%%%%%%%%%%%%5
In this section we study if it is possible to establish a connection between null geodesics and QNFs in the eikonal approximation by computing the Lyapunov exponent.
Firstly, in order to study the unstable null geodesics in the equatorial plane ($\theta=\pi/2$), we use the standard Lagrangian formalism \cite{Chandrasekhar}, so that, the corresponding Lagrangian associated with the line element (\ref{Ernstmetric}) reads
\begin{equation}
\mathcal{L}=\frac{1}{2}\left(\Lambda^{2}\left(f(r)\dot t^{2}-\frac{\dot r^{2}}{f(r)}\right)-\frac{r^{2}}{\Lambda^{2}}\dot \phi^{2}\right)\,.
\end{equation}
Therefore, from this Lagrangian  the generalized momenta are
\begin{align}
    p_t&=\Lambda(r)^{2}f(r) \dot t=E\,,\\
  p_\phi&=\frac{r^{2}}{\Lambda(r)^{2}}\dot\phi=L\,, \\
  p_r&=\frac{\Lambda(r)^{2}}{f(r)}\dot r\,.
\end{align}
Note that the Lagrangian is independent of both $t$ and $\phi$, so $p_t$ and $p_\phi$ are two integrals of motion. Also, the Hamiltonian is given by
\begin{equation}
2\mathcal{H}=2(p_{t}\dot t-(p_{r}\dot r+p_{\phi}\dot\phi)-\mathcal{L}
=\frac{E^{2}}{\Lambda^{2} f(r)}-\frac{L^{2}\Lambda^{2}}{r^{2}}-\frac{\Lambda^{2}\dot r^{2}}{f(r)}=0\,.
\end{equation}
Thus, from the above equation we have:
\begin{equation}\label{potencial}
    V_r(r)=\frac{E^{2}}{\Lambda^{4}}-\frac{f(r)L^{2}}{r^{2}}\,,
\end{equation}
where we have used $V_{r}=\dot r^{2}$.
Now, in order to obtain the circular geodesics, two conditions
must be satisfied, namely $V_{r}(r_c) =V'_{r}(r_c) =0$, which yields 
\begin{align}
 \frac{E^{2}}{L^{2}} &=\frac{f(r_c)\Lambda(r_c)^{4}}{r_c^{2}}\,, \\
 L^{2}&=\frac{4 r_c^{3}E^{2}\Lambda'(r_c)}{\Lambda^{5}(r_c)(2 f(r_c)-r_c f'(r_c))}\,.
\end{align}
As was pointed out in \cite{Dhurandhar,Lim:2015oha}, the condition  for the existence of  two possible radii outer the event horizon $r=2M$ is that $MB< 0.09468$. The inner radius is unstable an the outer one is stable.
In order to obtain the unstable radius that satisfies
\begin{equation}\label{fotonic}
3 r_{c}^3B^{2}-5Mr_{c}^{2}B^{2}-r_{c}+3 M=0\,,
\end{equation}
we consider for convenience the dimensionless units $x_c=\frac{r_c}{M}$ and $\beta=BM$. Also,
we expand the radius of circular geodesics in terms of small magnetic field B in the following form
\begin{equation}\label{expand}
   r_{c}= r_{0}+r_{1}\beta+r_{2} \beta^{2}\,.
\end{equation}
 Now, by replacing Eq. (\ref{expand}) into (\ref{fotonic}), we obtain
\begin{equation}
    x_{c}=3+36 B^2 M^{2}\,.
\end{equation}
Note a small correction to the Schwarzschild  radius of circular geodesics due to the magnetic field. The unstable circular geodesics possesses a larger radius than that of the Schwarzschild $x_c=3$.\\

On the other hand, it is known that the relation between QNMs and unstable circular null geodesics in the eikonal limit can be established for some spacetimes. For that we will use  the WKB method because it gives the correct approximation of QNMs in the eikonal limit.
Here, the central wave equation is given by (\ref{schrodinger}), and the effective potential in the eikonal limit $\ell \rightarrow \infty $ takes the form:
\begin{equation}
   V(r)\approx f(r)\left(\frac{\ell^{2}}{2 r^{2}}+4 m^{2}B^{2}\right)\,, 
\end{equation}
where the magnetic terms will be always small in comparison with the centrifuge terms, then the obtained effect is a small correction to Schwarzschild metric. The maximum value of the potential is found at $r_{0}=3M+\frac{36 B^{2} m^{2} M^{3}}{\ell^{2}}$. 
The QNMs leads to the following form
\begin{equation}
    \frac{Q_{0}(r_0)}{\sqrt{2 Q_{0}''(r_0)}}=i\left(n+\frac{1}{2}\right)\,,
\end{equation}
where $Q_{0}=\omega^{2}-V(r)$, $Q_0''\equiv \frac{d^{2}Q_{0}}{dr_{*}^{2}}$.
Following the Ref. 
\cite{Cardoso:2008bp}, where the authors showed that angular velocity $\Omega_c$ at the unstable null geodesic and the Lyapunov exponent $\lambda$, determining the instability timescale of the orbit agree with analytic WKB approximations for QNMs
\begin{equation}
\label{omegas}
\omega_{QNM}=\ell \Omega_{c}-i(n+\frac{1}{2})|\lambda|\,,
\end{equation}
where $\Omega_{c}=\Lambda^{2}(r_{c})\sqrt{\frac{f(r_{c})}{r_{c}^{2}}}$, $\lambda=\frac{\Lambda(r_{c})}{\sqrt{2}}\sqrt{\frac{f(r_{c})r_{c}^{2}V^{\prime \prime}_{r}(r_{c})}{L^{2}}}$, and 
\begin{equation}
V^{\prime \prime}_{r}(r_{c})=\frac{L^{2}}{r_{c}^{2}}\left(-6\frac{f(r_{c})}{r_{c}^{2}} +4\frac{f^{\prime}(r_{c})}{r_{c}}-f^{\prime \prime}(r_{c})-4f(r_{c})\frac{\Lambda{\prime\prime}(r_{c})}{\Lambda(r_{c})}\right)\,.
\end{equation}
So, by using Eq. (\ref{omegas}), we obtain 
\begin{equation}\label{frecuencias}
 \omega_{QNM} =\ell (1+B^{2}r_{c}^{2})^{2}\sqrt{\frac{r_c-2M}{r_{c}^{3}}}-i\left(n+\frac{1}{2}\right)\frac{1+B^{2}r_c^{2}}{r_c^{2}}\left((r_c-2M)(3(r_c-4 M)-4r_c^{2}B^{2}(r_c-2 M))\right)^{\frac{1}{2}}\,.
\end{equation}
Clearly, the QNMs in the eikonal limit are modified by the presence of the magnetic field $B$. Note that, the above equation for $m=\pm \ell$, $r_c= 3M+ 36 B^{2} M^{3}$, 
can be written as 
\begin{equation}
 \omega_{QNM} =\ell \left(\frac{1}{3\sqrt{3}M}+\frac{6 M B^{2}}{\sqrt{3}}\right)-i\left(n+\frac{1}{2}\right)\left(\frac{1}{3\sqrt{3}M}-\frac{5}{\sqrt{3}}M B^{2}\right)\,,   
\end{equation}
that matches with Eq. (\ref{omegarelation}) for the same parameters. In this way, at the equatorial plane, and with $m=\pm \ell$, there is a connection between the null geodesics and QNMs in the eikonal limit. The reason for this, is due to the position of the maximum of the potential given by (\ref{radius}) converges to the radius of circular geodesics when $m = \pm \ell$ in the eikonal limit. It is possible to think about the coincidence from another point of view; it is well known that in classical mechanics a particle orbiting on the equatorial plane and immersed in an uniform magnetic field perpendicular to that plane will have an orbital angular momentum $L=L_{z}$. Now, if we want to recover the classical situation from quantum mechanics, first we should recall that the orbital angular momentum vector is quantized, its magnitude is given by  $L^2=\ell(\ell+1)\hbar^2$ and the projection along the z-axes is $L_{z}=m\hbar$; therefore, in the particular case when $m= \ell $ or $m=-\ell$ and taking the limit  $\ell\rightarrow\infty$ it is possible to recover the classical setting.
It is worth noting that for $B=0$ the Schwarzschild QNMs in the optic geometric limit are recovered \cite{Cardoso:2008bp}.

%%%%%%%%%%%%%%%%%%%%%%%%%%%%%%%%%%%5

\section{Final remarks}
\label{conclusion}

In this work, we studied the propagation of charged massive scalar fields in the background of four-dimensional Ernst black holes. Then, by using the WKB method, we showed that  there is a critical scalar field mass, i.e, for a scalar field mass less than a critical mass the decay rate of the QNMs decreases when the harmonic angular number $\ell$ increases; and for a scalar field mass greater than the critical mass the behaviour is inverted, i.e,  the longest-lived modes are always the ones with lowest angular number recovering the standard behaviour. Additionally, we showed that there is a critical external magnetic field
that exhibits the same behaviour with respect to the angular harmonic numbers, as well as, a critical scalar field charge. 
On the other hand, concerning to the QNFs, for small values of $\ell$, we showed that for negative values of $m$,  
the real oscillation frequency grows, and the damping rate is decreasing when the magnetic field is increasing, whereas for positive values of $m$ 
the real oscillation frequency decreases very slightly and then begin to grow, and the damping rate increases and then decreases when the magnetic field increases.\\

On the other hand, we have shown that the spacetime allows stable quasibound states and the spectrum splits in $2 \ell+1$ branches, and the separation between the branches increases with the magnetic field, that is the analogous to the splitting of the energy levels of an atom in an external magnetic field, which is the well-known Zeeman effect. 
For the QBS, and negative values of $m$, 
the real oscillation frequency grows, and the damping rate is increasing when the magnetic field is increasing, whereas for positive values of $m$ 
the real oscillation frequency decreases very slightly and then begin to grow, and the damping rate increases and then decreases when the magnetic field  increases. So, the real oscillation frequency of the QNMs and QBS has a similar behaviour, whereas for the damping rate the behaviour is opposite, when the magnetic field is increasing.
Finally, we have shown that the unstable null geodesic in the equatorial plane is connected with the QNMs for $m = \pm \ell$ when $\ell \rightarrow \infty$.

\acknowledgments
This work is partially supported by ANID Chile through FONDECYT Grant Nº 1220871 (P.A.G., and Y. V.).  P.A.G. would like to thank the Facultad de Ciencias, Universidad de La Serena for its hospitality. R.B. would like to thank the Facultad de Ingenier\'{i}a y Ciencias, Universidad Diego Portales for its hospitality.

\end{document}